**Enhancement of ultrafast demagnetization rate and Gilbert damping driven by femtosecond laser-induced spin currents in $Fe_{81}Ga_{19}/Ir_{20}Mn_{80}$ bilayers**

Wei Zhang[1,2], Qian Liu[3], Zhe Yuan[3], Ke Xia[3], Wei He[1], Qing-feng Zhan[4], Xiang-qun Zhang[1], and Zhao-hua Cheng[1,2,5*]

[1]State Key Laboratory of Magnetism and Beijing National Laboratory for Condensed Matter Physics, Institute of Physics, Chinese Academy of Sciences, Beijing 100190, China

[2]School of Physical Sciences, University of Chinese Academy of Sciences, Beijing 100049, China

[3]The Center for Advanced Quantum Studies and Department of Physics, Beijing Normal University, 100875 China

[4]State Key Laboratory of Precision Spectroscopy, School of Physics and Materials Science, East China Normal University, Shanghai 200241, China

[5]Songshan Lake Materials Laboratory, Dongguan, Guangdong 523808, China

**Abstract**

In spintronics applications, ultrafast spin dynamics have to be controlled at femtosecond (fs) timescales via fs-laser radiation. At such ultrafast timescales, the effect of the Gilbert damping factor α on ultrafast demagnetization time $\tau_M$ should be considered. In previous explorations for the relationship between these two parameters, it was found that the theoretical calculations based on the local spin-flip scattering model do not agree with the experimental results. Here, we find that in $Fe_{81}Ga_{19}$(FeGa)/$Ir_{20}Mn_{80}$(IrMn) bilayers, the unconventional IrMn thickness dependence of α results from the competition between spin currents pumped from the ferromagnetic (FM) FeGa layer to the antiferromagnetic (AFM) IrMn layer and those pumped from the AFM layer to the FM layer. More importantly, we establish a proportional relationship between the change of the ultrafast demagnetization rate and the enhancement of Gilbert damping induced by the spin currents via interfacial spin chemical potential $\mu_s$. Our work builds a bridge to connect the ultrafast demagnetization time and Gilbert damping in ultrafast photo-induced spin currents dominated systems, which not only explains the disagreement between experimental and theoretical results in the relation of $\tau_M$ with α, but provides further insight into ultrafast spin dynamics as well.

*To whom all correspondence should be addressed. zhcheng@iphy.ac.cn

# I. INTRODUCTION

The understanding of spin dynamics from nanosecond (ns) down to femtosecond (fs) timescales is an essential task towards the realization of ultrafast spintronic devices in the frequency range from GHz to THz [1,2]. The study of ultrafast demagnetization time, $\tau_M$, is one of the most challenging problems in laser-induced ultrafast spin dynamics. The Gilbert damping factor, $\alpha$, is of the utmost importance for high frequency switching of spintronic devices. Since both $\tau_M$ and $\alpha$ require a transfer of angular momentum from the electronic system to the lattice, the unification of these two seemingly unrelated parameters can facilitate the exploration of the microscopic mechanism of laser-induced ultrafast spin dynamics. An inversely proportional relationship between $\tau_M$ and $\alpha$ was predicted by theoretical calculations based on the local phonon-mediated Elliott-Yafet scattering mechanism [3-5] as well as the stochastic Landau-Lifshitz-Bloch (LLB) model [6]. However, the relationship between $\tau_M$ and $\alpha$ has been debated for over one decade [7]. Until now, all experimental results have shown that $\tau_M$ increases with $\alpha$ [8-12].

Apart from the local spin-flip scattering mechanism [13], we proposed that the non-local spin currents should be taken into account to coordinate the contradiction in the relationship between $\tau_M$ and $\alpha$. Previous work suggested that the superdiffusive spin current contributed to ultrafast demagnetization [14], whilst the Gilbert damping could also be enhanced via non-local spin currents in ferromagnetic (FM)/nonmagnetic (NM) [15] and FM/antiferromagnetic (AFM) heterostructures [16]. Femtosecond laser irradiation of ferromagnetic thin films is a fascinating novel approach to create large spin currents [17,18]. Figure 1(a) shows that in the case of time-resolved magneto-optical Kerr effect (TRMOKE) experiments, hot electrons excited by fs-laser pulses can travel at high velocities and over tens of nanometers through the films. The difference of mean free path between spin majority and spin minority hot electrons in ferromagnetic thin films generates superdiffusive spin currents on fs timescales. Such spin currents dissipated at the interface of the heterostructure result in the out-of-equilibrium spin accumulation represented by spin chemical potential $\mu_s$. Moreover, figure 1(b) shows the damped magnetization precession around the effective field could be influenced via spin current. Tveten *et al.* [19] predicted that the ultrafast demagnetization time $\tau_M$ could be described in the language of spin current-induced damping $\alpha_{sp}$ in magnetic heterostructures based on the electron-magnon scattering theory. However, the experimental evidence on the connection of ultrafast demagnetization time with damping driven by fs laser-induced spin currents is not yet understood.

## II. RESULTS

### A. Sample properties

$Ir_{20}Mn_{80}$ ($t_{IrMn}$)/$Fe_{81}Ga_{19}$ (10 nm) bilayers [20] were deposited on optically transparent single-crystalline MgO (001) substrates in a magnetron sputtering system with a base pressure below $3\times10^{-7}$ Torr. The substrates were annealed at 700 °C for 1 h in a vacuum chamber and then held at 250 °C during deposition. FeGa layers were obliquely deposited at an incidence angle of 45°. The IrMn layers were deposited while continuously rotating the substrates. In order to induce an exchange bias (EB) along the FeGa [010] direction, a magnetic field of 500 Oe provided by a permanent magnet was applied along the MgO [110] axis during growth. After deposition, a 3 nm protective Ta layer was deposited on the samples to avoid oxidation. The static longitudinal Kerr loops of $Fe_{81}Ga_{19}$ (10 nm)/$Ir_{20}Mn_{80}$ ($t_{IrMn}$) along FeGa [010] direction with various AFM IrMn thicknesses ($t_{IrMn}$) at room temperature were acquired using a laser diode with a wavelength of 650 nm.

Figure 2(a) shows the longitudinal Kerr loops of $Fe_{81}Ga_{19}$ (10 nm)/$Ir_{20}Mn_{80}$ ($t_{IrMn}$ nm) along FeGa [010] direction with various AFM IrMn thicknesses ($t_{IrMn}$) at room temperature, whereas the thickness of FM FeGa layer was fixed at 10 nm. For $t_{IrMn} \leq 2nm$, the width of the hysteresis loops is enlarged with no obvious shift along the *x*-axis, implying that the thickness of IrMn layer is too thin to form an antiferromagnetic order for pinning the magnetization reversal of FeGa [21] (Insert in Fig. 2(b) (left)). For $t_{IrMn} > 2nm$, the antiferromagnetic orders are well established, and consequently the antiferromagnetic moments pin FM ones reversal to induce a unidirectional anisotropy (Insert in Fig. 2(b) (Right)). The loops therefore exhibit evidently exchange bias behavior. The exchange bias field achieves a value of about 60 Oe when $t_{IrMn} > 2nm$, whilst the largest value of coercivity (~72 Oe) occurs at $t_{IrMn} =$ 2 nm.

**B. TRMOKE measurements for ultrafast demagnetization and Gilbert damping**

We performed the polar TRMOKE experiment to measure ultrafast demagnetization time under a saturated applied field of 20 kOe in the normal direction of the samples [22]. The details of the TRMOKE experiment are described in **APPENDIX A.** Figure 3(a) shows the demagnetization curves for various IrMn thicknesses with a maximum magnetization quenching of ~10% [23,24]. The temporal changes of the Kerr signals $\Delta\theta_k(t)$ were normalized by the saturation value $\theta_k$ just before the pump laser excitation. The time evolution of magnetization on sub-picosecond timescales can be fitted according to Eq. (1) in terms of the three-temperature model (3TM) [17].

$$-\frac{\Delta M(t)}{M} = \left\{\left\{\left[\frac{A_1}{(t/\tau_0+1)^{0.5}} - \frac{A_2\tau_E - A_1\tau_M}{\tau_E - \tau_M}e^{-\frac{t}{\tau_M}} - \frac{\tau_E(A_1-A_2)}{\tau_E - \tau_M}e^{-\frac{t}{\tau_E}}\right]\Theta(t)\right\} * G(t,\tau_G)\right\} * G(t,\tau_G) \quad (1)$$

where $*G(t,\tau_G)$ represents the convolution product with the Gaussian laser pulse profile, $\tau_G$ is the full width at half maximum (FWHM) of the laser pulses, $\Theta(t)$ is a step function, $\delta(t)$ is the Dirac delta function. $A_1$ represents the value of $\frac{\Delta M(t)}{M}$ after equilibrium between electrons, spins, and lattices. $A_2$ is proportional to the initial electrons temperature rise. Here, we used the 780 nm laser as the pump pulse to excite the magnetic system out of equilibrium, while the 390 nm laser pulse was used as a probe beam. Therefore, in Eq. (1), the state filling effects during pump probe experiment are neglected due to the different wavelength of pump and probe beams used in this study. The cooling time by heat diffusion is described by $\tau_0$, which should be about one order of magnitude larger than $\tau_E$ representing the timescale of electron-phonon interactions. The best-fitted value of $\tau_E = 500\ fs$ for all samples is in good agreement with that of previous reports [18]. The fitting parameters in Eq. (1) are shown in Table I, from which one notes the pulse width is 350 fs for all the samples. In our experimental setup, the time-resolution is about 80 fs. In order to obtain a high time resolution, we measured the ultrafast demagnetization with very fine step of time delay (15 fs). The values of ultrafast demagnetization time (120-220 fs) obtained from Eq. (1) are defined as the time needed for the magnetization to reach a level of $e^{-1}$ of its maximum demagnetization. The time needed for magnetization to reach its maximum demagnetization (>500fs) should be longer than the time extracted from Eq. (1). A similar result was reported by B. Vodungbo *et al*.[25]. The very large temporal stretching of the laser pulse up to 430 fs was attributed to the conversion of the incident laser pulse into a cascade of hot electrons. This could be one of the possible reasons resulting in the spread of laser pulse up on the samples in this study. Via changing the single parameter, $\tau_M$, we can accurately reproduce the experimental results for various samples. The ultrafast demagnetization time $\tau_M$ was observed to decrease from $220 \pm 10$ fs for $t_{IrMn}$ = 0 nm to $120 \pm 10$ fs for $t_{IrMn}$ = 2 nm, then increase with further increasing $t_{IrMn}$ [Fig. 3(b)].

The precessional frequency and damping factor can be derived by means of the TRMOKE signals as well [26, 27]. Figure 4(a) shows the typical time evolution of the polar component of magnetization after pump laser excitation at different fields applied along with the [110] direction of FeGa for $t_{IrMn} = 2\ nm$. It is observed clearly that the spin precession process can be influenced obviously by applied fields. The exact values for $f$ with various applied fields can be obtained using the damped harmonic function added to an exponential-decaying background:

$$\Delta M(t) = A + Bexp(-vt) + Cexp\left(-\frac{t}{\tau}\right)sin(2\pi ft + \varphi) \qquad (2)$$

where $A$ and $B$ are the background magnitudes, and $v$ is the background recovery rate. $C, \tau, f$ and $\varphi$ are the magnetization precession amplitude, relaxation time, frequency and phase, respectively. The field dependence of frequency $f$ extracted from the fitting procedure is shown in Fig. 4(b). We note that the experimental $f$-$H$ relation can be reproduced very well by Kittel equation (3) [27].

$$\left(\frac{2\pi f}{\gamma}\right)^2 = \frac{1}{M_s^2}H_1H_2 \qquad (3)$$

with $H_1 = -2K_{Out} + 4\pi M_s^2 + 2K_u\cos^2\varphi_M + 2K_1 - K_1\sin^2 2\varphi_M + HM_s\cos(\varphi_M - \varphi_H) + K_{eb}\cos\varphi_M$ and

$H_2 = 2K_1\cos4\varphi_M + 2K_u\cos2\varphi_M + M_sH\cos(\varphi_M - \varphi_H) + K_{eb}\cos\varphi_M$.

And $\gamma = \gamma_e g/2$ is the gyromagnetic ratio. $\varphi_M$ and $\varphi_H$ are the angles of in-plane equilibrium $M$ and $H$ respect to the FeGa [010] easy axis. $K_1, K_u, K_{eb}$ and $K_{Out}$ are the in-plane magnetocrystalline, uniaxial, unidirectional and out-of-plane magnetic anisotropy constants of FeGa films, respectively. The value of magnetocrystalline anisotropy constant is $K_1 = 4.5 \times 10^5\ erg/cm^3$ for the samples with various AFM layer thickness during the fitting procedure and the uniaxial magnetic anisotropy constant $K_u = (1.5 \pm 0.3) \times 10^5 erg/cm^3$. For $t_{IrMn} = 3\ nm$ and 5 nm, the unidirectional magnetic anisotropy constant of $K_{eb} = 3 \times 10^4 erg/cm^3$ has to be included for more accurate fitting, although it is one order magnitude smaller than those of magnetocrystalline and uniaxial anisotropy.

The effective Gilbert damping factor $\alpha_{eff}$ shown in Fig. 4(c) is determined from the relaxation time $\tau$ by Eq. (4) [28]:

$$\alpha_{eff} = 2/\tau\gamma(H_1 + H_2) \tag{4}$$

Since the overall effective damping factor $\alpha_{eff}$ consists of intrinsic damping and extrinsic damping whereby the second one arises from both the two-magnon-scattering and the dephasing effect in the samples, the overall effective Gilbert damping factor decreases monotonously to a constant value with increasing the applied field (Fig. 4(c)). As one of the mainly extrinsic contributions, the two-magnon-scattering induced damping has been extensively studied in exchange biased heterostructures [29-34]. The mature theory was developed to explain the two-magnon scattering process due to spatial fluctuations of anisotropy and exchange bias field [30,35]. The two-magnon scattering process comes from the scatterings of the uniform ($k = 0$) precession mode into nonuniform modes ($k \neq 0$ magnons) that are degenerate in frequency. This process is described by the Hamiltonian, in which the spatial fluctuation in the exchange coupling caused by interface roughness determines the scattering strength. The roughness gives rise to a large fluctuating field because the FM magnetization interacts alternatively with one or the other AF sublattice via the atomic exchange coupling. It is a well-known relaxation mechanism effective in exchange biased heterostructures due to the interface roughness occurring on the short length scales. When a low external field comparable with the exchange bias field was applied, the two-magnon scattering effect will result in the increase of Gilbert damping with the exchange bias field according to previous reports [33, 34]. However, as shown in Ref. 36, a strong enough applied field can be used to exclude the contributions from the two-magnon-scattering, where the value of Gilbert damping factor keeps as a constant with various two-magnon-scattering strength. Based on this result, a similar method using strong enough external fields was applied in this study to exclude the two-magnon-scattering effect. Moreover, previous works show that the two-magnon-scattering induced damping increases with precession frequency because of the increased degeneracy of spin waves [37, 38]. Our work demonstrated that the damping factor keeps almost a constant value at high enough applied fields, indicating the minor contributions from the two-magnon-scattering to Gilbert damping. Besides, it has been demonstrated previously that the two-magnon-scattering contributions decrease monotonously with increasing the film thickness [33, 34]. This again disagrees with the tendency of thickness dependence of damping at high applied field shown in Fig. 5(c). Therefore, in this study, the two-magnon-scattering strength was suppressed effectively by applying a high enough external field. On the other hand, inhomogeneities in FeGa thin film may cause variations in the local magnetic anisotropy field. It leads to the variations of spin orientations when the external field is not large enough, and gives rise to the enhanced damping arising from spin dephasing effect [28]. However, an applied field (~ kOe) much larger than the anisotropy field makes the spin orientation uniform, as a result, the dephasing effect was suppressed largely. Based on the above analysis, the intrinsic part of damping is independent of the external field or precession frequency, while the extrinsic part including both the dephasing effect and the two-magnon-scattering effect are field-dependent. In order to avoid the effect of the extrinsic damping factor, the intrinsic damping factors were obtained by fitting the overall damping factor as the function of applied fields with the Eq. (5) [39, 40] shown as the red line in Fig. 4(c):

$$\alpha_{eff} = \alpha + \alpha_1 e^{-H/H_0} \tag{5}$$

where $\alpha$ and $\alpha_1 e^{-H/H_0}$ are the intrinsic and extrinsic parts of the damping factor, respectively.

For the derivation of spin precessional frequency as well as the Gilbert damping, the similar producers as shown above were adapted to various samples. Figure 5(a) shows the precessional frequency from oscillation curves with various IrMn thicknesses. Since the exchange bias field and coercivity are much weaker than applied fields, the *f-H* curves of FeGa films are therefore slightly different with various AFM layer thicknesses, which is in contrast to the observation that the enhanced uniaxial anisotropy of Fe/CoO bilayers [28] increases the precessional frequency largely. More importantly, we find the effective damping factor $\alpha_{eff}$ decreases with applied fields [Fig. 5(b)]. The solid lines represent the fitting expression shown as the Eq. (5). Interestingly, the effective Gilbert damping factors drop to a nearly constant value as the intrinsic damping factor when the applied fields increase strong enough to suppress the extrinsic contributions as stated above.

The values of the intrinsic damping factor as a function of the thickness of the IrMn layer are illustrated in Fig. 5(c). It increases firstly and reaches the maximum value when the thickness of the IrMn layer at $t_{IrMn}$ = 2 nm, and finally decreases with further increasing the thickness of the IrMn AFM layer. A drastic change of 2.5 times for damping occurs at $t_{IrMn}$ = 2 nm. Similarly, S. Azzawi et al. showed around 2 times enhancement of damping in NiFe/Pt bilayers when a continuous Pt capping layer is just forming at 0.6 nm by TRMOKE measurements [41]. Moreover, once a continuous IrMn layer is forming at 2 nm, the accompanied strong intrinsic anisotropy of AFM would contribute partly to the damping enhancement superimposed to spin pumping effect. This has been demonstrated previously by W. Zhang et al where the damping of Py/IrMn bilayers is 3 or 4 times larger than that in the Py/Cu/IrMn samples [42]. Based on the discussions in Fig. 4, we can exclude the extrinsic mechanisms such as the two-magnon-scattering and the dephasing effect as the dominant contributions to the damping process when the external fields are high enough [43]. Besides, FeGa alloys are particularly interesting because of their magneto-elastic properties [44]. The acoustic waves are possible to be triggered by ultrashort laser and as a result, spin precession would be excited non-thermally via a magnetoelastic effect [45]. However, this effect can be excluded based on the following reasons: firstly, the external field has to be applied along with the hard axis of FeGa, otherwise, the magnetization precession cannot be induced. It agrees with the fact that the canted magnetization from the easy axis is necessary when the spin precession arising from instantaneous anisotropy change accompanied by ultrafast demagnetization occurs [26]. In contrast, the occurrence of spin precession from the magnetoelastic effect is independent of initial magnetization orientation. Secondly, in order to check the contribution of resonance mode from the magnetoelastic effect, we performed a fast Fourier transform in **APPENDIX B**. Only the uniform field-dependent precession mode was excited at present study. This is not the expected behavior for the acoustically induced modulation of the magneto-optical effects. Therefore, the magnetoelastic effect of FeGa was suppressed largely in this study. It probably because the laser fluence of around 1 mJ/cm$^2$ is not high enough to induce a large amplitude of strain pulse. According to Ref. 45, the oscillations amplitude of acoustic mode increases linearly with the laser energy density within the probed range. Moreover, the FeGa material with a thickness as thick as 60 nm is preferred to induce an obvious magnetoelastic behavior [46], while 10 nm at the present experiment is probably too thin. As a result, the intrinsic damping can be influenced by the following paramenters: (1) magnetocrystalline anisotropy of FM [47]; (2) exchange bias field [30, 31, 36], and (3) spin pumping effect at the interface between FM and the AFM [15, 16, 42, 48]. In the case of FeGa/IrMn bilayers, the magnetocrystalline anisotropy constant of FeGa $K_1 = 4.5 \times 10^5 erg/cm^3$, which is obtained from Fig. 4 and Fig. 5, is invariant with AFM layer thickness. Moreover, referring to Fig. 2(b), it seems that there is no direct relationship between the intrinsic damping factor and the exchange bias field $H_{eb}$. When the applied field is far higher than the exchange bias field, both the precessional frequency and the damping factor show independence of exchange bias field [36]. Therefore, the IrMn thickness dependence of the intrinsic damping is not attributed to the magnetocrystalline anisotropy and the exchange bias field. Due to the strong spin-orbit coupling of the heavy metal (HM) Ir in the IrMn alloy, the contribution of spin pumping to the damping factor must be taken into account. It is noteworthy that the IrMn thickness dependence of damping in FeGa/IrMn is different from that in other normal FM/HM bilayers, where the damping factor increases monotonically with the thickness of HM layer and approaches a saturation value [49]. However, the damping of FeGa ferromagnetic layer decreases again after reaching a peak value at $t_{IrMn} = 2\ nm$. The change of the damping factor is always accompanied by the spin currents transfer between FM and AFM layers. More spin currents absorbed by the neighboring layer result in larger damping in the FM layer. An unconventional decrease of the damping factor implies that not only the effect of heavy metal Ir in IrMn alloy has to be taken into account, but also the antiferromagnetic magnetization. The heavy metal Ir serves as a perfect spin sink to absorb the spin currents, and consequently increases the damping in FeGa, while the antiferromagnetic magnetization in IrMn serves as a new source to compensate the dissipation of magnetization precession and decrease the damping of FeGa.

### C. First-principle calculations for IrMn layer thickness dependence of Gilbert damping

To understand the behavior of the IrMn thickness-dependent damping factor, we calculated the damping factor using the scattering theory of magnetization dissipation combined with the first-principles electronic structure [50]. The calculated FM/AFM bilayer structure shown in Fig. 6(a) are the same as that in the experiment. Here, the magnetic moments of AFM sublattices serve as not only a spin sink to absorb the spin current pumped from the adjacent FM layer, but also a spin current emitter to partly cancel the spin pumping effect of the FM. The interfacial exchange coupling forces the magnetic moments of the IrMn sublattices in a few layers near the interface to precess following the adjacent FM, generating spin currents back into the FM layer [Fig. 6(b)]. Based on this model, the enhancement of damping due to the spin current $\alpha_{sp} = \Delta\alpha = \alpha_{t_{IrMn}} - \alpha_{t_{IrMn}=0nm}$ as a function of IrMn thickness was calculated and shown as the solid circle in Fig. 6(c). It increases firstly to a peak value at $t_{IrMn}$ = 2 nm, and then drops with further increasing the IrMn layer thickness. When $t_{IrMn} \leq 2$ nm, the thickness of the IrMn layer is too thin to establish the antiferromagnetic order, which can be supported by the negligible exchange bias as shown in Fig. 2(b). In this case, the pumped spin current from the AFM back into the FM to partially cancel the spin pumping effect by the FM is largely reduced because of the disorder of the antiferromagnetic moments as illustrated on the left side in Fig. 6(b). In this region, therefore, the magnetic moments in the AFM serve as a perfect spin sink to absorb the spin current pumped from the adjacent FM resulting in a significant enhancement in the damping factor. For the samples with the thickness of IrMn $t_{IrMn} > 2nm$, however, the antiferromagnetic order is well established and the accompanied exchange bias is remarkably large (See Fig. 2(b) and its insert). Because of the exchange coupling between FM and AFM at the interface, the magnetic moments of the AFM sublattices in a few layers near the interface is forced to precess following the magnetic moment of the FM, while those far away from the interface would keep static. Such an exchange spring effect at the interface caused spin precession in the AFM layer, and consequently, spin currents would be transferred from AFM to the FM layer. Moreover, these spin currents from the AFM would be enhanced due to the coherent precession of magnetization in different sublattices as illustrated in the right side of Fig. 6(b). The exchange spring effect induced precession of the AFM has two effects: (1) the AFM has intrinsic damping that increases the overall damping of the FM/AFM bilayer. (2) the precessional motion of magnetic moments in AFM sublattices pumps spin currents into the FM, which cancels partly the spin pumping by the FM. As a result, the overall damping of the bilayers is reduced. From the solid circles in Fig. 6(c), one can find that the damping decreases with increasing $t_{IrMn}$ when $t_{IrMn} > 2$ nm, indicating that the latter effect of the pumped spin currents is dominant over the intrinsic damping. Besides, by comparing the calculated and experimental values [Fig. 6(c) and (d)], one can find that the calculated Gilbert damping is larger than the experimental one for $t_{IrMn} = 1$ nm. The reason for the deviation is the assumption of a perfectly flat FeGa/IrMn interface in the calculation, which leads to a larger spin current pumped from the FM. Unfortunately, it is almost impossible to fabricate the perfectly flat film when the thickness is less than 1 nm.

In order to separate the contribution of the precession of the magnetic moment of the AFM sublattice to damping, we also calculated the damping by assuming perfectly static AFM ordered IrMn without precession (solid diamonds in Fig. 6(c)) and a paramagnetic IrMn layer with vanishing Néel order (solid triangles in Fig. 6(c)). The calculated results demonstrate that if the magnetic moments of the AFM sublattice either do not precess or align randomly, the IrMn layers serve only as a perfect spin sink to absorb the spin currents pumped from the adjacent FM resulting in a significant enhancement of damping. The damping increases monotonically to a saturation value with IrMn thickness, which is similar to that of heavy metals [49].

### D. Relationship between ultrafast demagnetization rate and Gilbert damping induced by non-local spin currents

The central strategy of our study is to establish a direct correlation between $\tau_M$ and α. According to Fig. 3(b) and Fig. 5(c), we find that the femtosecond laser-induced ultrafast demagnetization time $\tau_M$ and the Gilbert damping α show an opposite IrMn thickness dependence in FeGa/IrMn bilayers. By plotting $\tau_M$ versus α as shown in Fig. 7(a), one can clearly observe that the value of $\tau_M$ decreases with α, suggesting that spin transport plays an additional dissipation channel for accelerating the ultrafast demagnetization and enhancing the damping. The damping factor $\alpha_{t_{IrMn}}$ for $t_{IrMn} > 0$ nm is ascribed to the spin pumping effect induced by various AFM thicknesses $\alpha_{sp}$ and the contribution from the FM itself $\alpha_{t_{IrMn}=0nm}$. To give further insight into the relationship, we replotted Fig. 7(a) by using the change of the ultrafast demagnetization rate $\Delta \frac{1}{\tau_M} = \frac{1}{\tau_M}|_{t_{IrMn}} - \frac{1}{\tau_M}|_{t_{IrMn}=0nm}$ versus the enhancement of Gilbert damping $\alpha_{sp} = \Delta\alpha = \alpha_{t_{IrMn}} - \alpha_{t_{IrMn}=0nm}$ induced by the spin current. An approximately linear relationship is confirmed and shown in Fig. 7(b), which can be fitted using Eq. (6): (For the derivation of Eq. (6), please see **APPENDIX D** for details)

$$\Delta \frac{1}{\tau_M} = \frac{\mu_s}{\hbar} \Delta\alpha, \tag{6}$$

Where $\Delta \frac{1}{\tau_M}$, $\Delta\alpha$ represents the enhancement of ultrafast demagnetization rate and Gilbert damping induced by the spin current, respectively, $\mu_s$ is the spin chemical potential, and $\hbar$ is the Planck constant. A reasonable value of $\mu_s \approx 1\ eV$ which is similar to that of spin splitting in 3d transition metals was obtained by the linear fitting using Eq. (6).

The spin chemical potential $\mu_s$ is proportional to spin accumulations at the interface between different layers. It contributes largely to ultrafast demagnetization according to the model of laser-induced ultrafast superdiffusive spin transport in layered heterostructures [14, 51]. There is a large difference in velocities or lifetimes for spin-dependent hot electrons [52]. As a result, the transport properties of hot electrons are spin-dependent. For instance, the minority-spin electrons excited by ultrashort laser survive for a quite short time and they decay to non-mobile bands approximately at the position they were excited. Instead, majority-spin electrons have longer lifetimes and higher velocities. So they leave fast from the excitation region after being created, resulting in part of the demagnetization process. Because the directions of motion for all the electrons are random, they can obtain a velocity directed back towards the ferromagnetic film. A second part of the demagnetization is ascribed to the backflow of spin-minority electrons from the substrate or the neighbor layer. Spin-majority electrons entering the ferromagnetic layer will find good transport properties and continue diffusing without severely decaying. However, spin-minority electrons experience a considerable worsening of the transport properties as soon as they enter the ferromagnetic layer. The consequence is that they are trapped at the entrance of the ferromagnetic layer, giving rise to the spin accumulations at the interface. Nevertheless, the quantitative description for spin accumulations during ultrashort laser-induced demagnetization in heterostructures is still lacking. This work aims at filling this gap by relating ultrafast demagnetization time and Gilbert damping. A detailed calculation for the value of 1 eV for spin chemical potential obtained in this experiment is highly desirable.

The non-local spin currents dissipated at the interface of FeGa/IrMn open an additional channel to accelerate the ultrafast demagnetization and enhance the Gilbert damping. However, in the case of the sample with $t_{IrMn}$ = 0 nm without the assistant AFM layer, both the local spin-flip and non-local spin transport mechanisms probably contribute to the ultrafast demagnetization in the ferromagnetic layer. For instance, based on the breathing Fermi-surface model of the Gilbert damping and the Elliott-Yafet relation for the spin-relaxation time, a relation shown as Eq. (7) is established between the conductivity-like Gilbert damping $\alpha$ and ultrafast demagnetization time $\tau_M$ [10].

$$\tau_M = \frac{M}{\gamma F_{el} p b^2} \alpha \tag{7}$$

Taking the values of $\tau_M|_{t_{IrMn}=0nm}$ and $\alpha|_{t_{IrMn}=0nm}$ are 220 fs and 0.004, respectively, a value of $\alpha/\tau_M = 1.8 \times 10^{10} s^{-1}$ is derived. This value is reasonable and agrees well with that of 3d transition metal Ni calculated by the breathing Fermi-surface model [53], indicating that the ultrafast demagnetization of ferromagnetic FeGa film itself is mainly governed by the local spin-flip scattering events. Nonetheless, we have to address that, ultrafast demagnetization in the ferromagnetic layer was accelerated and the Gilbert damping was enhanced via the interfacial spin accumulations once the IrMn layer was attached.

### III. CONCLUSIONS

The unconventional IrMn thickness dependence of α is attributed to the cancellation of the spin currents pumped from the AFM IrMn layer to the FM FeGa layer. We establish a proportional relationship between the change of ultrafast demagnetization rate and the enhancement of Gilbert damping induced by the spin currents via interfacial spin chemical potential. This result can facilitate the utilization of ultrafast spintronic devices in the THz region.

**Acknowledgments**

This work is supported by the National Key Research Program of China (Grant Nos. 2015CB921403, 2016YFA0300701, and 2017YFB0702702), the National Natural Sciences Foundation of China (Grant Nos. 91622126, 51427801, and 51671212) and the Key Research Program of Frontier Sciences, CAS (Grant Nos. QYZDJ-SSW-JSC023, KJZD-SW-M01 and ZDYZ2012-2). The work at Beijing Normal University is partly supported by the National Natural Sciences Foundation of China (Grant Nos. 61774017, 61704018, and 11734004), the Recruitment Program of Global Youth Experts and the Fundamental Research Funds for the Central Universities (Grant No. 2018EYT03). The work at East China Normal University is partly supported by the National Natural Sciences Foundation of China (Grant No. 11874150).

**APPENDIX A: TRMOKE MEASUREMENTS**

In this study, the dynamical process of fast and ultrafast spin dynamics was measured by TRMOKE. The experiments were carried out using an all-optical pump-probe technique. A train of optical pulses with a wavelength of 780 nm, 55 fs duration, and 100 nJ/pulse is generated at 5.2 MHz repetition rate by a Ti: sapphire oscillator (FEMTOLASER, XL-100). A 200 µm thickness BBO crystal was used to double the frequency of femtosecond laser. The laser beam from the source is split into both 780 nm and 390 nm beams. We use the 780 nm laser as the pump pulse to excite the magnetic system out of equilibrium, while the 390 nm laser pulse was used as a probe beam to measure the subsequent magnetization dynamics with the timescale from sub-picosecond to nanosecond. The pump laser beam is much stronger than the probe with an intensity ratio of about 100 for all the measurements. Both the pump and probe beams are incident along the normal axis (z-axis) of the samples. The detection geometry is only sensitive to the out-of-plane component of the magnetization $M_z$. For fast spin dynamics, we applied various external fields along the $Fe_{81}Ga_{19}$ [110] direction to trigger the spin precession, while a large enough field about 20 kOe was applied along the $Fe_{81}Ga_{19}$ [001] direction to obtain the ultrafast demagnetization curves. We adjusted the pump laser fluence from 1 mJ/cm$^2$ to 1.25 mJ/cm$^2$ to obtain the same maximum quenching for various samples. The pump and probe beams are focused onto the samples with spot diameters of ~10 µm and ~5 µm via an objective lens, respectively. For the spin precession measurements, the scheme of the TRMOKE experiment is illustrated in Fig. 8. The signals are sensitive with the polar component of magnetization after pump laser excitation at different fields applied along the [110] direction of FeGa.

**APPENDIX B: FAST FOURIER TRANSFORM ANALYSIS**

The ferromagnetic FeGa is a famous material for its magneto-elastic properties. After femtosecond laser irradiation, an external field-independent resonance mode would be triggered due to the excitation of coherent acoustic phonons. However, only one field-dependent resonance mode was excited in this study according to fast Fourier transform analysis in Fig. 9.

**APPENDIX C: FIRST-PRINCIPLE CALCULATIONS**

The electronic structure of FeGa/IrMn bilayer is calculated self-consistently using the local density approximation of the density functional theory. The spin-dependent potentials, charge and spin densities are obtained with the minimal basis of tight-binding linear muffin-tin orbitals. In the calculation of the total damping, the scattering region consisting of the repeated FeGa/IrMn bilayers are connected to two semi-infinite Cu leads. We have introduced the thermal lattice disorder into a 4x4 supercell and displaced the atoms in the scattering region randomly away from their equilibrium positions with a Gaussian distribution. The root-mean-square atomic displacements of the Gaussian distribution are determined using a simple Debye model with the Debye temperature of 470 K. The two-dimensional Brillouin zone of the supercell is sampled by a 24x24 k-mesh corresponding to the 96x96 mesh for the Brillouin zone for the 1x1 unit cell. The effect of magnons in the FM FeGa is neglected in our calculation. This is because the magnetic damping is dominated by electrons at the Fermi level in metals, which can efficiently transfer spin angular momentum into the orbital motion via spin-orbit interaction. In metals and alloys, the influence of magnon-phonon coupling is negligible except for near the Curie temperature [54].

If magnetization precession occurs only in the FM FeGa layer, the calculated damping enhancement does not sensitively depend on the specific order of the AFM IrMn. Here we take two limits: the perfectly antiferromagnetic ordered IrMn and the paramagnetic IrMn (the magnetic moments of Mn are randomly distributed such that both the Néel order and total magnetization vanish). The damping enhancements calculated for the two cases are nearly identical, where the damping factor is enhanced and saturates at the thickness of 2 nm. It indicates that the pumped spin current by the precessional FeGa is immediately absorbed by the IrMn layer. The large moment on the Mn atom can absorb the pumped transverse spin current efficiently. On the other hand, the AFM IrMn is forced to process due to the interfacial exchange coupling, however, the efficient of the spin current generation by AFM depends on its specific order. It is suppressed largely in the case of paramagnetic IrMn because of the cancellation via magnetic moments with various orientations shown on the left side of Fig. 6(b) in the main text. In contrast, the efficient of the spin current generation by the AFM is enhanced remarkably by the coherent precession of the ordered magnetic moments shown in the right side of Fig. 6(b) in the main text. And the cone angle of precessional IrMn is modeled to exponentially decay from the interface with a typical decay length of 2 nm. The precessional AFM has mainly two contributions to the damping enhancement of the bilayer. First, the AFM has intrinsic damping that increases the total energy loss during the magnetization dynamics. The second effect is that the precessional AFM pumps spin current into the FM that cancels partly the spin pumping by the FM and decreases the damping enhancement.

## APPENDIX D: DERIVATION OF EQ. (6) IN THE MAIN TEXT

It is well known that the magnetic moment $\vec{M}_s$ is proportional to the spin angular momentum $\vec{S}$ via gyromagnetic ratio $\gamma = \frac{g\mu_B}{\hbar}$ :

$$\vec{M}_s = \gamma \vec{S} \tag{8}$$

where $g$ is Lande factor, $\mu_B$ is Bohr magneton. Normally, we take $\vec{m} = V\vec{M}_s$ as the total magnetic moments, where $V$ is the volume of the atom.

$\tau_M$ is the ultrafast demagnetization time. Therefore, the value of $\frac{1}{\tau_M}$ is taken as the demagnetization rate. The demagnetization is always accompanied by the dissipation of spin angular momentum, and hence the rate of spin angular momentum dissipation is :

$$\frac{\vec{m}}{\gamma} \bullet \frac{1}{\tau_M}. \tag{9}$$

On the other hand, the spin current $\vec{j^s}$ of per unit area generated by spin pumping effect

reads:

$$\vec{j^s} = \frac{1}{4\pi} g_{eff} \vec{\mu_s}, \tag{10}$$

where $g_{eff}$ is the effective interfacial spin-mixing conductance including the influence of the backflow spin current from the AFM IrMn to FeGa, $\vec{\mu_s}$ is the spin accumulation-driven chemical potential. The pumped spin current across the interface is $\vec{I^s} = \vec{j^s}A$, where $A$ is the area of the interface.

$$g_{eff} = \frac{4\pi M_s d \Delta\alpha}{g\mu_B}, \tag{11}$$

where $d$ is the thickness of the ferromagnetic layer, $\Delta\alpha = \alpha_{t_{IrMn}} - \alpha_{t_{IrMn}=0nm}$ is the enhancement of Gilbert damping induced by the absorption and generation of spin current via various IrMn thicknesses.

Therefore, if we correlate the spin angular momentum dissipated by the ultrafast demagnetization and that induced by spin pumping, the relationship reads:

$$\frac{\vec{m}}{\gamma} \bullet \frac{1}{\tau_M} = \vec{I}^s \tag{12}$$

And then we take Eq. (10) into Eq. (12), we can correlate the parameters $\tau_M$ and $\alpha$ via:

$$\frac{1}{\tau_M} = \frac{\mu_s}{\hbar} \Delta\alpha, \tag{13}$$

To exclude the contributions from local spin-flip scattering mechanisms to the ultrafast demagnetization rate represented by $\frac{1}{\tau_M}\big|_{t_{IrMn}=0nm}$, the value of $\frac{1}{\tau_M}$ is replaced by

$$\Delta\frac{1}{\tau_M} = \frac{1}{\tau_M}\bigg|_{t_{IrMn}} - \frac{1}{\tau_M}\bigg|_{t_{IrMn}=0nm}.$$

**Figure Captions**

Table. I. Values of the main fit parameters of ultrafast demagnetizations curves for various thicknesses of the samples.

FIG.1. (color online) Basic concept of both ultrafast demagnetization and spin precession induced by spin currents. (a) The excitation of fs laser pulse transforms slow majority-spin d electrons (red) into fast sp electrons, thereby launching a spin current towards the AFM layer. The spin current crossing the interface results in the spin accumulation at the interface represented by spin chemical potential $\mu_s$. (b) The typical time evolution of magnetization after femtosecond laser irradiation measured by TRMOKE experiment.

FIG. 2. (Color online) Static magnetic properties of of MgO/Fe$_{81}$Ga$_{19}$ (10 nm)/Ir$_{20}$Mn$_{80}$ (t nm) bilayers. (a) Longitudinal-MOKE loops with various thicknesses of IrMn layer $t_{IrMn}$. (b) Coercivity $H_c$ and exchange bias field $H_{eb}$ as a function of IrMn layer thickness $t_{IrMn}$.

FIG.3. (Color online) Ultrafast demagnetization. (a) Ultrafast demagnetization curves with various IrMn layer thicknesses. The solid lines represent the fitting results by Eq. (1) in the text. The insert shows the configuration of the measurement for ultrafast

demagnetization. (b) Ultrafast demagnetization time as a function of IrMn layer thickness.

FIG.4. (Color Online) Spin precession. (a) TRMOKE signals of FeGa/IrMn bilayers with $t_{IrMn}=2$ nm in various applied fields. (b) Precessional frequency as a function of applied fields. (c) Effective Gilbert damping constant as a function of applied fields.

FIG. 5. (Color Online) Frequency and damping of spin precession. (a) Frequency of spin precession as a function of applied fields with various IrMn thickness. The solid lines represent the fitting results by Kittle equations. (b) Effective Gilbert damping constants as a function of applied fields with various IrMn thicknesses. (c) Intrinsic Gilbert damping as a function of IrMn thickness.

FIG.6. (Color online) Results of First-principle calculations. (a) Illustration of the ferromagnet (FM)/antiferromagnet (AFM) structure employed to investigate the spin transport. (b) The configuration of the IrMn magnetic moments located at the first layer near the interface. (c) The calculated damping enhancement as a function of the thickness of the antiferromagnetic IrMn. The solid circles show the calculated damping enhancement with the precession of AFM magnetic moments. The solid diamonds show the calculated damping enhancement with perfectly static AFM ordered IrMn without precession, while the solid triangles correspond to the calculated values using a static paramagnetic IrMn layer with vanishing Néel order. (d) The experimental damping enhancement as a function of the thickness of antiferromagnetic IrMn.

FIG.7. (Color online) (a) Ultrafast demagnetization time as a function of Gilbert damping. (b) The variation of ultrafast demagnetization rate as a function of Gilbert damping enhancement. The red line indicates the fitting via Eq. (6) in the text.

FIG. 8. (Color online) Scheme of TRMOKE experiment for spin precession dynamics.

FIG. 9. (Color online) Fourier transform spectra measured between 0.85 kOe and 3.0 kOe for $t_{IrMn} = 2\ nm$.

Table I

| $t_{IrMn}$ (nm) | $\tau_M$ (fs) | $\tau_E$ (fs) | $\tau_0$ (ps) | $\tau_G$ (fs) | A1 | A2 |
|---|---|---|---|---|---|---|
| 0 | 220±10 | 500 | 5 | 350 | 0.8 | 2 |
| 1 | 160±10 | 500 | 6 | 350 | 0.8 | 2 |
| 2 | 120±10 | 500 | 7 | 350 | 0.8 | 2 |
| 3 | 145±10 | 500 | 4 | 350 | 0.8 | 2 |
| 5 | 200±10 | 500 | 5 | 350 | 0.8 | 2 |

.

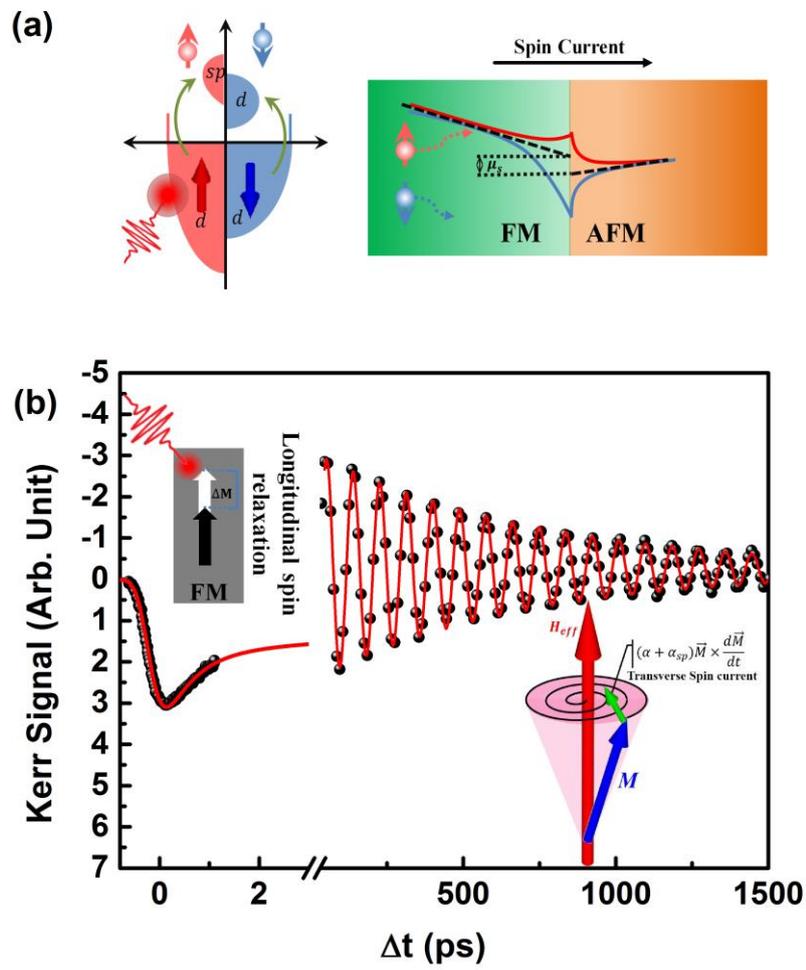

FIG.1.

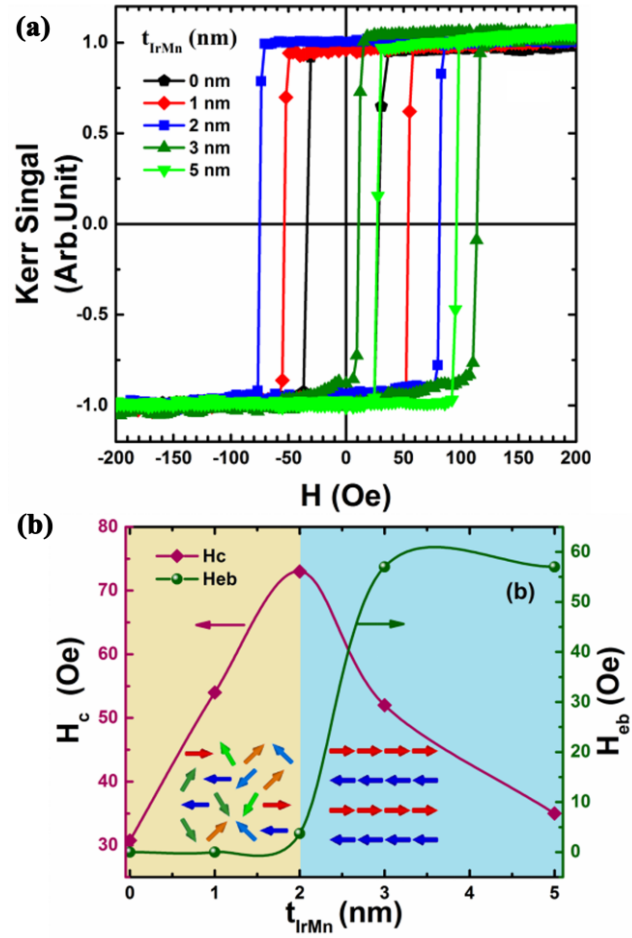

FIG. 2.

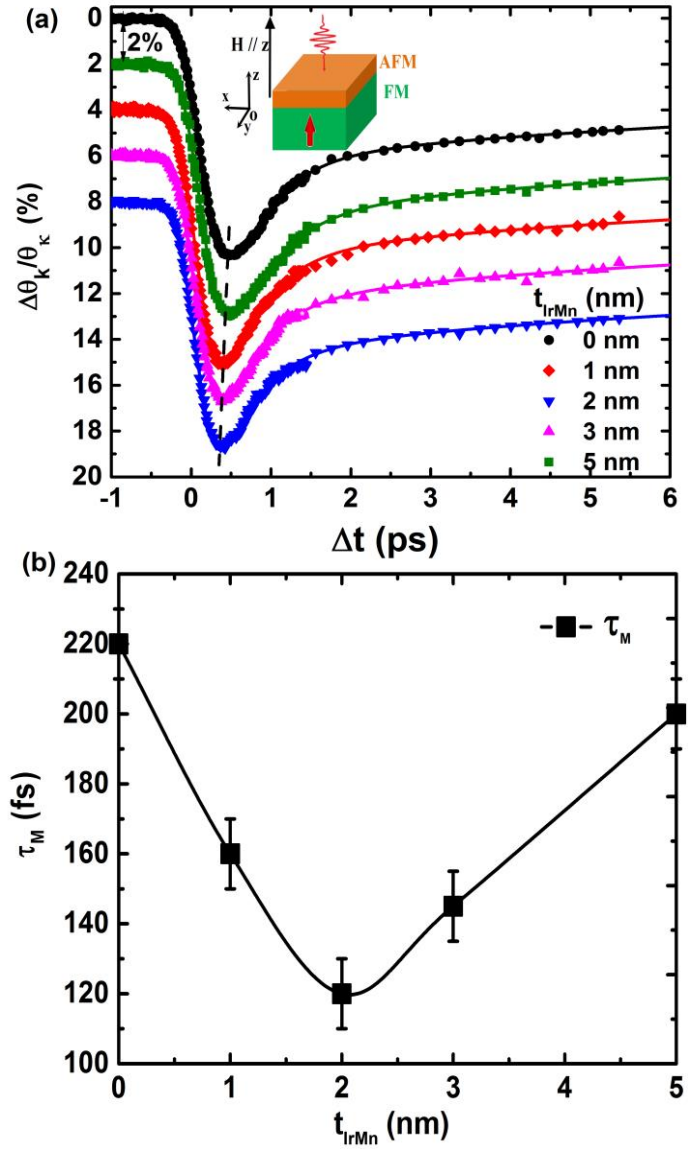

FIG.3.

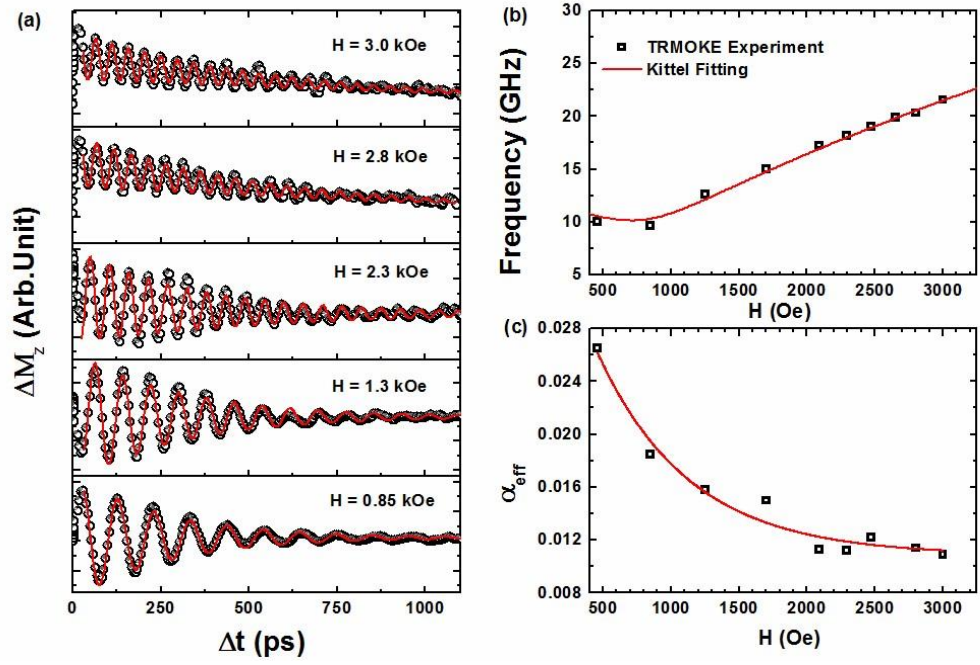

FIG.4.

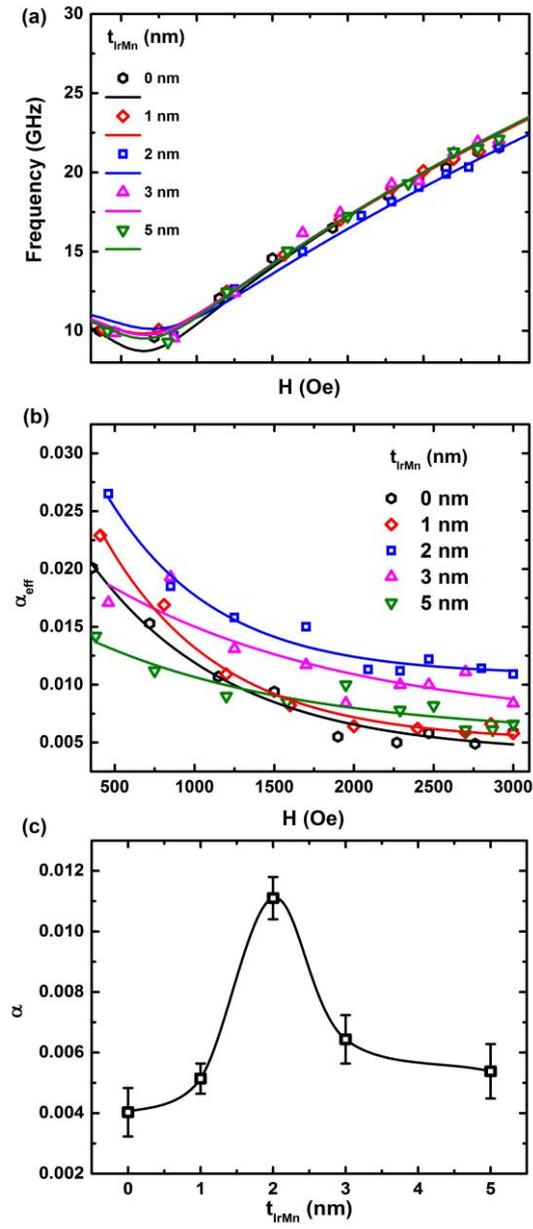

FIG. 5.

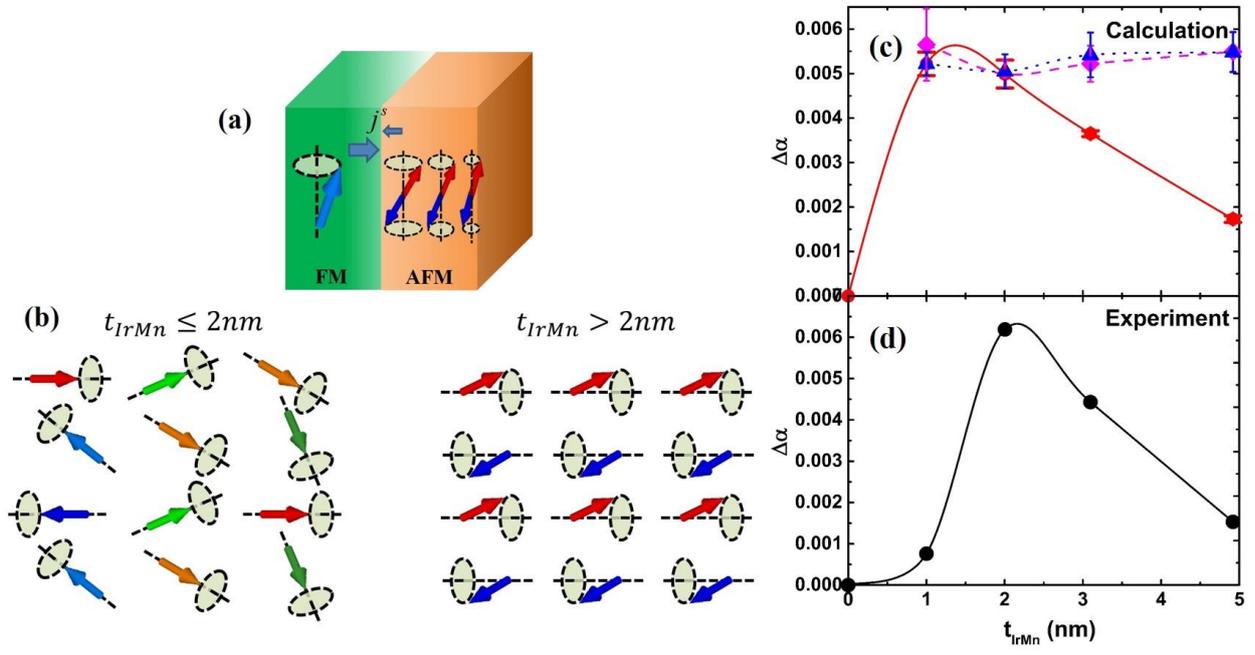

FIG.6.

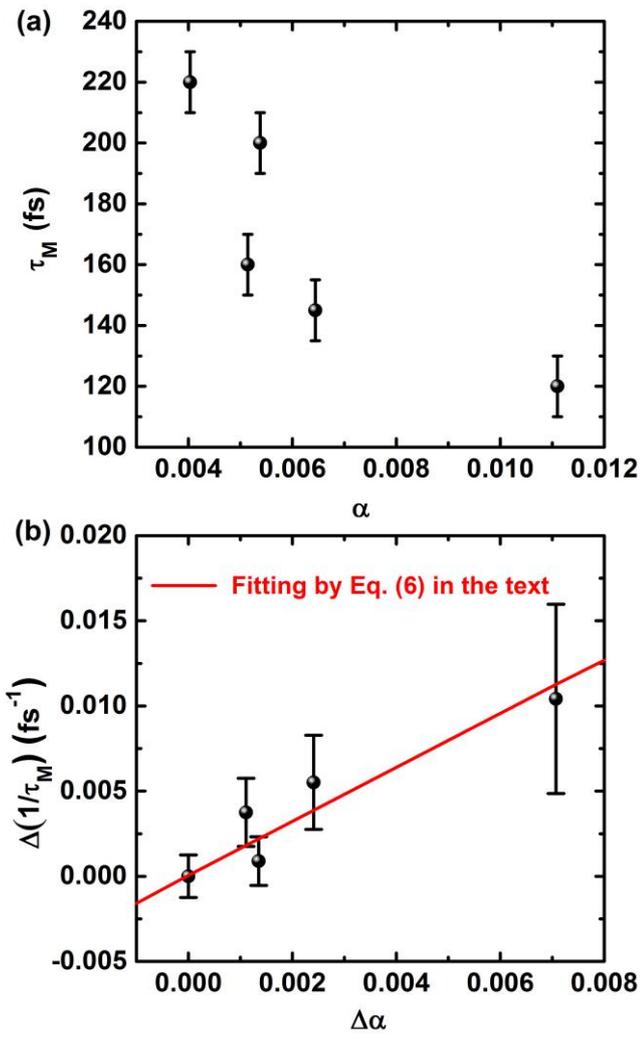

FIG.7.

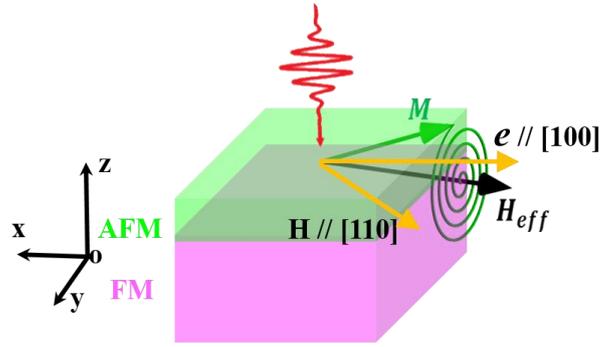

FIG. 8.

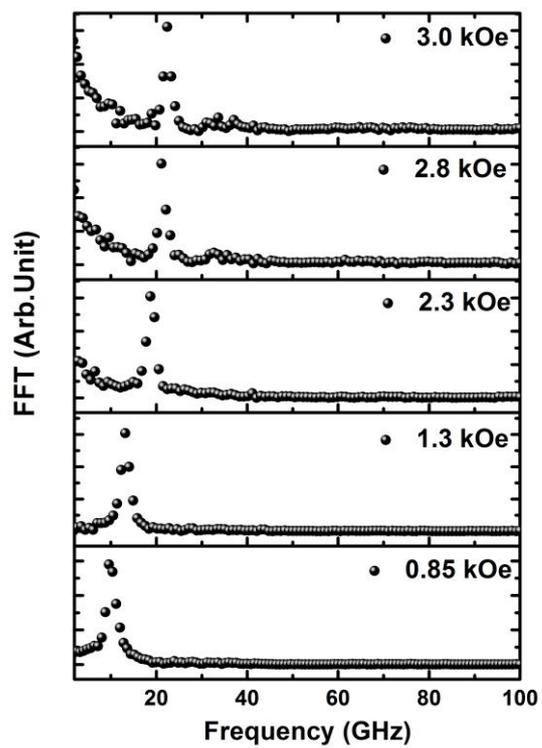

FIG.9.

44